\documentclass[12pt,tightenlines,showpacs,showkeys]{revtex4-1}
\usepackage{graphicx,color,amssymb}

\begin{document}

\title{Centaurus A as a point source of Ultra-High Energy Cosmic Rays}
\author{Hang Bae Kim}\email{hbkim@hanyang.ac.kr}
\affiliation{Department of Physics and The Research Institute of
Natural Science, Hanyang University, Seoul 133-791, Korea}

\begin{abstract}
We probe the possibility
that Centaurus A (Cen A) is a point source of
ultra-high energy cosmic rays (UHECR) observed by PAO,
through the statistical analysis of the arrival direction distribution.
For this purpose, we set up the Cen A dominance model for the UHECR sources,
in which Cen A contributes the fraction $f_{\rm C}$ of the whole UHECR
with energy above $5.5\times10^{19}\,{\rm eV}$ and
the isotropic background contributes the remaining $1-f_{\rm C}$ fraction.
The effect of the intergalactic magnetic fields on the bending
of the trajectory of Cen A originated UHECR
is parameterized by the gaussian smearing angle $\theta_s$.
For the statistical analysis,
we adopted the correlational angular distance distribution (CADD)
for the reduction of the arrival direction distribution,
and the Kuiper test to compare the observed and the expected CADDs.
Using CADD, we identify the excess of UHECR in the Cen A direction
and fit the CADD of the observed PAO data by varying two parameters
$f_{\rm C}$ and $\theta_s$ of the Cen A dominance model.
The best-fit parameter values are $f_{\rm C}\approx0.1$
(The corresponding Cen A fraction observed at PAO is
$f_{\rm C,PAO}\approx0.15$, that is, about 10 out of 69 UHECR.)
and $\theta_s=5^\circ$ with the maximum probability $P_{\rm max}=0.29$.
Considering the uncertainty concerning the assumption of isotropic background
in the Cen A dominance model, we extend the viable parameter ranges to
the $2\sigma$ band,
$0.09\lesssim f_{\rm C,PAO}\lesssim 0.25$ and
$0^\circ\lesssim \theta_s\lesssim 20^\circ$.
This result supports the existence of a point source
extended by the intergalactic magnetic fields in the direction of Cen A.
If Cen A is actually the source responsible for the observed excess of UHECR,
the average deflection angle of the excess UHECR implies
the order of $10\,{\rm nG}$ intergalactic magnetic field
in the vicinity of Cen A.
\end{abstract}
\pacs{98.70.Sa}
\keywords{ultra high energy cosmic rays, Centaurus A, active galactic nuclei}

\maketitle

\section{Introduction}

The origin of the ultra-high energy cosmic rays (UHECR) is a long-standing
puzzle \cite{Nagano:2000ve}.
The recent confirmation of the Greisen-Zatsepin-Kuzmin (GZK) suppression
in the cosmic ray energy spectrum
\cite{Abraham:2008ru,Abbasi:2007sv,AbuZayyad:2012ru}
implies that UHECR with energies above the GZK cutoff,
$E_{\rm GZK}\sim4\times10^{19}\;{\rm eV}$, mostly come from relatively close
extragalactic sources within the GZK radius $r_{\rm GZK}\sim100\,{\rm Mpc}$.
Furthermore, UHECR with these energies are expected not to be strongly affected
by the galactic or extragalactic magnetic field, so that their arrival
directions keep some correlation with the source distribution and can be used
to trace the sources of UHECR.
Recently Pierre Auger Observatory (PAO) released the updated UHECR data with
energy $E\ge5.5\times10^{19}\;{\rm eV}$ \cite{Abreu:2010ab}.
These data can be used for tracing the distribution of UHECR sources
through the statistical analysis of the arrival direction distribution.
Beginning with the report of the correlation between UHECR and active galactic
nuclei (AGN) by the PAO collaboration
\cite{Cronin:2007zz,Abraham:2007si,Abreu:2010ab},
several attempts was made to identify the UHECR sources.
Most attempts aimed to test the plausibility of a certain kind of high energy
astrophysical objects whose number ranges from a few to a few hundred
\cite{Takami:2009bv,Cuesta:2009ap,Koers:2008ba,Takami:2008ri,Kashti:2008bw,
Abbasi:2008md,Abbasi:2005qy}.
In this paper, we aim to test the other extreme possibility that
a single dominating source is responsible for the large part of observed UHECR.

The most mentioned candidates for a strong UHECR source are
Centaurus A (Cen A) and Messier 87 in Virgo cluster,
which are active galaxies very close to us.
Many people suggested that Cen A might be the source of UHECR
\cite{Romero:1995tn,Anchordoqui:2001nt,Isola:2001ng,Hardcastle:2008jw,
Honda:2009xd,GopalKrishna:2010wp,Fraija:2010ip,Anchordoqui:2011ks}.
Cen A is located in the southern sky,
near the center of the exposure region of the PAO experiment.
Therefore, the PAO data provide the best chance for checking
whether Cen A is a strong source of UHECR.
Actually, the correlation of UHECR with AGN was strengthened by many
UHECR observed around Cen A.
On the contrary, the number of UHECR around M87 is smaller
than the expected one considering its distance,
weakening the claimed AGN correlation.
Our purpose is to quantify the Cen A contribution in the observed UHECR
by PAO and thus try to establish the existence of a point source
in the Cen A direction.

For a single dominating source to be able to explain the large portion of
UHECR data, the intergalactic magnetic fields must play a significant role,
spreading UHECR over the large region of the sky around the source.
Unfortunately, our knowledge about the intergalactic magnetic fields is
rather poor yet. Modeling the intergalactic magnetic fields and testing it
has its own uncertainty.
For this reason, we choose a different and simpler strategy.
We parameterize the effect of intergalactic magnetic fields on the deflection
of UHECR trajectory by the gaussian spreading of UHECR arrival directions
around the sources.
In addition, we also need to consider the contribution from other sources
to explain the whole set of observed UHECR arrival directions.
For this purpose, we introduce one more parameter measuring the fraction of
Cen A contribution to observed UHECR.
Then we search for the values of two parameters with which the observed
UHECR arrival directions can be plausibly explained
through the statistical comparison of the arrival direction distributions.

Statistical comparison of the arrival direction distributions can be done
in many different ways.
In Ref.~\cite{Kim:2010zb,Kim:2012en},
we developed the new statistical test methods,
whose basic idea is that the two-dimensional distribution of arrival
directions is reduced to the one-dimensional probability distributions,
which can be compared by using the well-known Kolmogorov-Smirnov test
or its variants.
We proposed a few reduced one-dimensional distributions
suitable for the test of correlation between the UHECR
arrival directions and the point sources.
Among them, we adopt the correlational angular distance distribution method
and the flux-exposure value distribution method.
These methods will be briefly described in Sec.~\ref{sec3}.

This paper is organized as follows.
In section \ref{sec2}, we explain the Cen A dominance model
for the UHECR sources and the details needed for the Monte-Carlo
simulations of UHECR arrival directions.
In section \ref{sec3}, we briefly introduce our statistical methods for
comparing two arrival direction distributions.
In Section \ref{sec4}, the results of our analysis are presented.
We give a few discussions on the results and conclude in section \ref{sec5}.

\section{The single dominating source model for UHECR}
\label{sec2}

We examine the plausibility of the idea that the Cen A is the dominant
sources of UHECR through the statistical analysis of the arrival direction
distribution of UHECR.
For more definite interpretation of our analysis,
we need to solidify the UHECR source model with the Cen A dominance
and the methods adopted for statistical analysis.
In this section, we describe the details of the Cen A dominance model
for UHECR sources.

Cen A is located at $\alpha=201.37^\circ$, $\delta=-43.02^\circ$
in the equatorial coordinates and
the distance is estimated to be $3-5\,{\rm Mpc}$ \cite{Harris:2009wj}.
We consider Cen A as a smeared point sources of UHECR.
This is mainly to incorporate the fact that the trajectories of UHECR can be
bent by intervening magnetic fields.
We may model the intervening galactic and extragalactic magnetic fields between
Cen A and the earth. But, it would involve large arbitrariness due to our lack
of knowledge on extragalactic magnetic fields.
Instead of such detailed modeling of magnetic fields,
we simply assume that Cen A has a gaussian flux distribution on the sky
with a certain angular width $\theta_s$,
so that the effect of magnetic fields is measured through the parameter
$\theta_s$, called the smearing angle.
Then the UHECR flux at a direction $\hat{\bf r}$ contributed by Cen A can be
written as
\begin{equation}
F_{\rm CA}(\hat{\bf r}) = \overline{F}_{\rm CA}
\frac{\exp\left[-\left(\theta_j(\hat{\bf r})/\theta_s\right)^2\right]}
{N(\theta_s)}\,,
\end{equation}
where $\overline{F}_{\rm CA}$ is the averaged flux of Cen A,
$\theta(\hat{\bf r})=\cos^{-1}(\hat{\bf r}\cdot\hat{\bf r}_{\rm CA})$
is the angle between the direction $\hat{\bf r}$ and Cen A, and
$N(\theta_s)=(1/4\pi)\int d\Omega\exp[-(\theta(\hat{\bf r})/\theta_s)^2]$
is the normalization of smearing function.
For small $\theta_s$, $N(\theta_s)\approx\theta_s^2/4$
and for large $\theta_s$, $N(\theta_s)\approx1$.
For the small smearing angle $\theta_s$,
the average deflection angle is
$\langle\theta\rangle\equiv\int\theta e^{-(\theta/\theta_s)^2}d\Omega/
\int e^{-(\theta/\theta_s)^2}d\Omega \approx\theta_s$,
and $\Delta\theta\equiv\sqrt{\langle\theta^2\rangle-\langle\theta\rangle^2}
\approx\theta_s/2$.

Though Cen A can be a dominant source of UHECR,
it is very unlikely that Cen A is the only source of UHECR.
We need to consider the contribution from other distributed sources.
We consider, for the sake of simplicity,
that a certain fraction of UHECR are originated from Cen A,
while the remaining fraction of them are from
the isotropically distributed background contributions.
Then, the expected flux at a give arrival direction $\hat{\bf r}$ is
given by the sum of two contributions,
\begin{equation}
F(\hat{\bf r}) = F_{\rm CA}(\hat{\bf r}) + F_{\rm ISO}.
\end{equation}
Now we define the Cen A fraction $f_C$ to be
\begin{equation}
f_C = \frac{\overline{F}_{\rm CA}}{\overline{F}_{\rm CA}+F_{\rm ISO}}.
\end{equation}
The UHECR flux can be written as
\begin{equation}
F(\hat{\bf r}) = f_C\overline{F}\,
\frac{\exp\left[-\left(\theta(\hat{\bf r})/\theta_s\right)^2\right]}{N(\theta_s)}
+(1-f_C)\overline{F},
\end{equation}
where $\overline{F}=\overline{F}_{\rm C}+F_{\rm ISO}$.
Out of three parameters $\overline{F}_{\rm C}$, $F_{\rm ISO}$, and $\theta_s$,
the Cen A fraction $f_C$ and the smearing angle $\theta_s$ are treated as the
free parameters of the model, while the average flux $\overline{F}$ is fixed
by the total number of UHECR events.

To do the simulation for the observed arrival directions of UHECR,
we also need to take into account the efficiency of the detector
as a function of the arrival direction.
It depends on the location and the characteristics of the detector array.
Here we consider only the geometric efficiency which is determined
by the location and the zenith angle cut of the detector array.
Then the exposure function $h(\hat{\bf r})$ depends only on the declination
$\delta$,
\begin{equation}
h(\delta) = \frac{1}{\pi}\left[ \sin\alpha_m\cos\lambda\cos\delta
	+\alpha_m\sin\lambda\sin\delta\right],
\end{equation}
where $\lambda$ is the latitude of the detector array,
$\theta_m$ is the zenith angle cut, and
\[
\alpha_m=\left\{\begin{array}{ll}
0,            & \hbox{for\ } \xi > 1, \\
\pi,          & \hbox{for\ } \xi < -1, \\
\cos^{-1}\xi, & \hbox{otherwise}
\end{array}\right.
\ \hbox{with}\
\xi=\frac{\cos\theta_m-\sin\lambda\sin\delta}{\cos\lambda\cos\delta}.
\]
The expected flux at the detector array is proportional to
$F(\hat{\bf r})h(\hat{\bf r})$.
We also note that the Cen A fraction $f_{\rm C}$ is the fraction of Cen A
contribution over the whole sky.  It is in general different from the fraction
of Cen A contribution within the sky covered by a given detector array,
because the latter is masked by the exposure function.
We denote the latter, e.g.\ for PAO, by $f_{\rm C,PAO}$
to distinguish it from the former.

\section{Statistical Comparison of Two Arrival Direction Distributions}
\label{sec3}

We now turn to the statistical methods
to measure the plausibility of the Cen A dominance model.
In Refs.~\cite{Kim:2010zb,Kim:2012en}, we developed the simple comparison
method for the UHECR arrival direction distributions,
where the two-dimensional UHECR arrival direction distributions on the sphere
is reduced to one-dimensional probability distributions of some sort,
so that they can be compared by using the standard Kolmogorov-Smirnov (KS)
test or its variants.
In this paper, we adopt the reduction methods
called the correlational angular distance distribution (CADD)
and the flux-exposure value distribution (FEVD).
For a detailed explanation on these methods,
see Refs.~\cite{Kim:2010zb,Kim:2012en}.
Here we present briefly the basic ideas of these distributions
and how to calculate the probability measuring the plausibility.

$\bullet$ {\it Correlational Angular Distance Distribution}
This is the distribution of the angular distances of all pairs
UHECR arrival directions and the point source directions:
\begin{equation}
\label{CADD}
\hbox{CADD : }
\left\{ \cos\theta_{ij'}\equiv \hat{\bf r}_i\cdot\hat{\bf r}'_j
\;|\; i=1,\dots,N;\; j=1,\dots,M \right\},
\end{equation}
where $\hat{\bf r}_i$ are the UHECR arrival directions,
$\hat{\bf r}'_j$ are the point source directions,
and $N$ and $M$ are their total numbers, respectively.

$\bullet$ {\it Flux Exposure Value Distribution}
At a given arrival direction, the expected flux value is the product of
the UHECR flux expected from the UHECR source model
and the exposure function of the detector at that direction.
FEVD is the distribution of expected flux values at UHECR arrival directions:
\begin{equation}
\label{FEVD}
\hbox{FEVD : }
\left\{ F_i\equiv F(\hat{\bf r}_i)h(\hat{\bf r}_i)
\;|\; i=1,\dots,N \right\},
\end{equation}
where $\hat{\bf r}_i$ are the UHECR arrival directions,
$N$ is the total numbers of UHECR,
$F(\hat{\bf r}_i)$ and $h(\hat{\bf r}_i)$ are the UHECR flux
and the exposure function, respectively.

From the given point source set and the UHECR arrival direction data set,
we can get CADD and FEVD.
Then, we can apply KS test or its variants such as Kuiper test and
Anderson-Darling test to compare two CADDs or two FEVDs,
one from the observed UHECR data set and the other from the expected
(simulated) UHECR data set from the model under consideration.
In this analysis, we use Kuiper (KP) test because
its sensitivity is found to be most appropriate for our purpose and
the probability function of its statistic is available in analytic form.
The KP test is based on the cumulative probability distribution (CPD),
$S_N(x)=\int^xp(x')dx'$ and the KP statistic is the sum of maximum
difference above and below two CPDs,
\begin{equation}
\label{KP-statistic}
D_{\rm KP}=\max_{x}\left[S_{N_1}(x)-S_{N_2}(x)\right]
+\max_{x}\left[S_{N_2}(x)-S_{N_1}(x)\right].
\end{equation}
From the KP statistic $D_{\rm KP}$,
the probability that CADD/FEVD of the observed data
is obtained from the model under consideration can be estimated
using the Monte-Carlo simulations in general.
For the KP statistic $D_{\rm KP}$,
when the data in the distribution are all independently sampled,
the following approximation formula is available:
\begin{equation}
\label{KP-P-formula}
P(D_{\rm KP}|N_e) = Q_{\rm KP}([\sqrt{N_e}+0.155+0.24/\sqrt{N_e}]D_{\rm KP}),
\end{equation}
where $Q_{\rm KP}(\lambda)=2\sum_{j=1}^\infty(4j^2\lambda^2-1)
e^{-2j^2\lambda^2}$
and $N_e=N_1N_2/(N_1+N_2)$ is the effective number of data.
For FEVD, the number of data in the distribution is
same as the number of UHECR data.
For CADD, the number of data in the distribution is
the number of UHECR data times the number of point sources.
Therefore, for a single source case,
the number of data in the distribution CADD and FEVD
is same as the number of UHECR data.
This means that the data in CADD and FEVD are all independent,
and we can use the formula (\ref{KP-P-formula}) for both CADD and FEVD
for a single source model.
Now, $N_1=N_{\rm O}$, the number of observed UHECR data
and $N_2=N_{\rm S}$, the number of mock UHECR data.
We can make the expected distribution more accurate by increasing the number
of mock data $N_2$.
In the limit $N_2\rightarrow\infty$, the effective number
of data is simply $N_e=N_{\rm O}$.

\section{The results of statistical tests}
\label{sec4}

\begin{figure}
\centerline{\includegraphics[width=100mm]{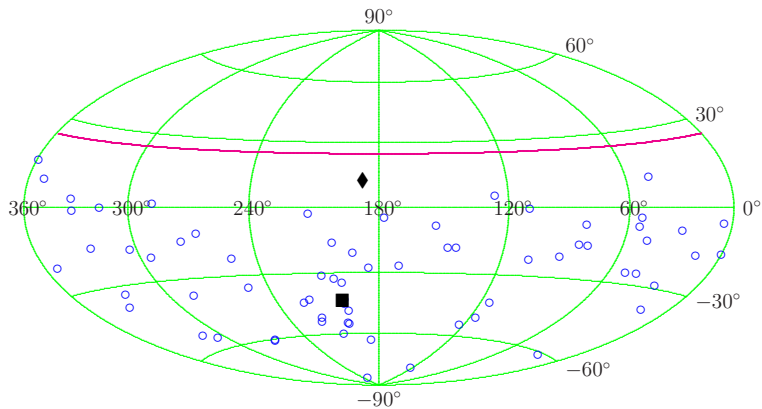}}
\caption{Distribution of the arrival directions of 69 UHECR,
represented by blue circles ({\color{blue}$\circ$}),
with energy $E\geq5.5\times10^{19}\,{\rm eV}$ reported by PAO in 2010,
in the equatorial coordinates plotted using the Hammer projection.
The solid red line represents the exposure boundary of the PAO experiment.
The locations of Centaurus A (${\color{black}\blacksquare}$) and
M87 (${\color{black}\blacklozenge}$) are also shown for reference.}
\label{skymap-data}
\end{figure}

For the observed UHECR data set,
we use the UHECR data released by PAO in 2010 \cite{Abreu:2010ab}.
The released data set contains 69 UHECR
with energy higher than $5.5\times10^{19}\,{\rm eV}$.
The PAO site has the latitude $\lambda=-35.20^\circ$
and the zenith angle cut of the released data is $\theta_m=60^\circ$.
The arrival direction distribution of the released PAO data is shown
in Fig.~\ref{skymap-data}.
The locations of Cen A and M87 are also marked for reference.
For the correlation test between the UHECR arrival directions and the
astrophysical objects, the choice of the energy cut can be crucial.
For low energy UHECR, the effects of intergalactic magnetic fields may be
so strong that the correlation of UHECR arrival directions
with their sources can be completely erased.
The energy cut $5.5\times10^{19}\,{\rm eV}$ of the released data was chosen
to be higher enough than the GZK cutoff so that their sources can be
restricted within the GZK radius $\sim 100\,{\rm Mpc}$.
This high value of the energy cut is good enough for correlation analysis.
Thus, we use the full set of released PAO data for our analysis.

\begin{figure}
\centerline{\includegraphics[width=100mm]{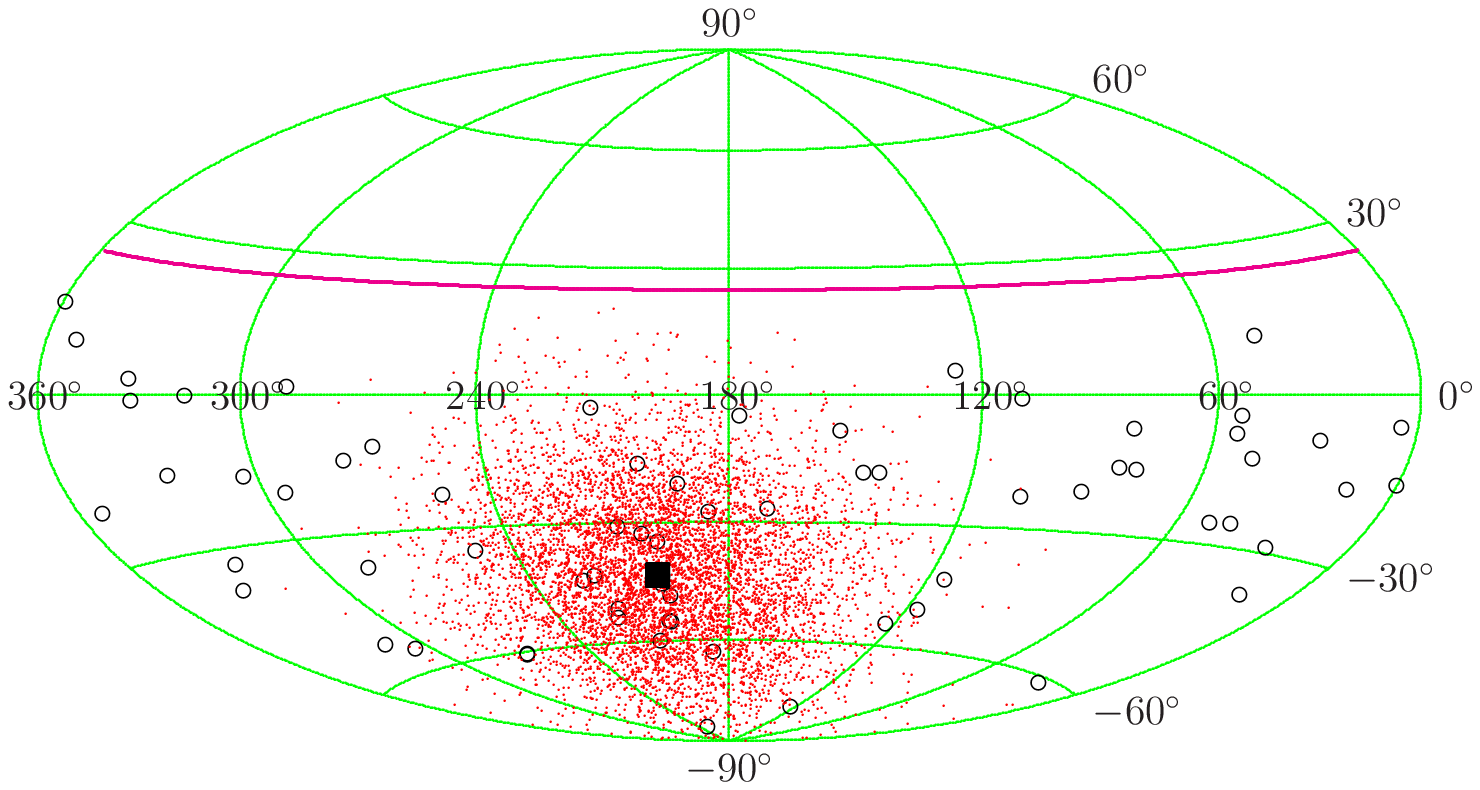}}
\centerline{\includegraphics[width=100mm]{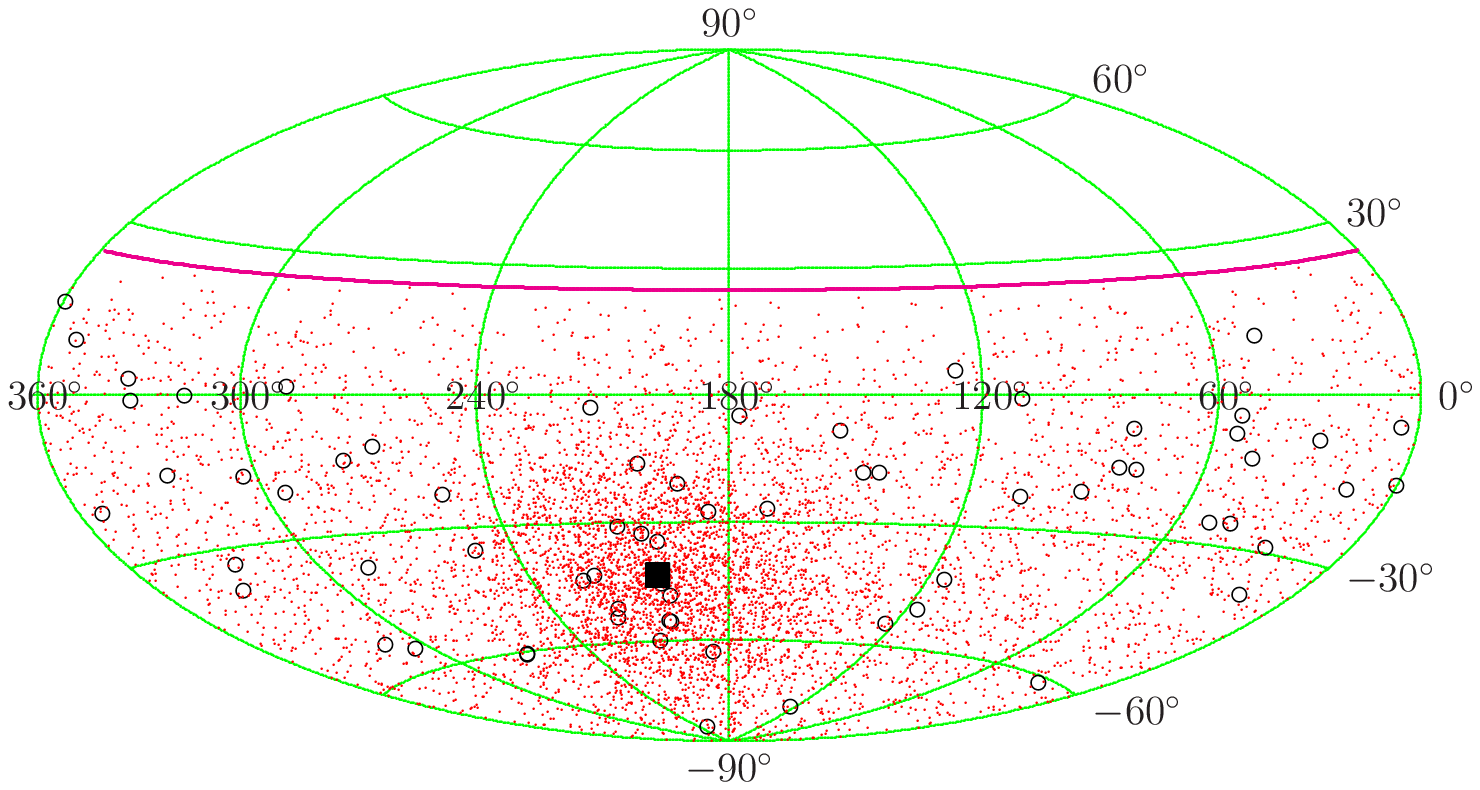}}
\caption{Distributions of the mock UHECR arrival directions (6900 events,
represented by small red dots) for the PAO experiment, obtained from the
Cen A dominance model for two different values of the Cen A fraction
$f_C=1.0$ and $f_C=0.3$
with the smearing angle $\theta_s=30^\circ$.}
\label{skymap-sim}
\end{figure}

From the Cen A dominance model,
the expected arrival direction distribution
can be obtained through the simulation.
As an illustration,
we show the arrival direction distributions of 6,900 mock UHECR
for two different values of the Cen A fraction, $f_C=1.0$ and $f_C=0.3$
with the same smearing angle $\theta_s=30^\circ$
in Fig.~\ref{skymap-sim}.
From the observed UHECR set and the mock UHECR set,
we obtain the observed and the expected CADDs/FEVDs
as described in Eqs.~(\ref{CADD}) and (\ref{FEVD}),
and calculate the KP statistic using Eq.~(\ref{KP-statistic}).
To obtain the KP statistic with sufficient accuracy,
we generate $10^6$ mock UHECR events.
Though we still have small fluctuations in the KP statistic
for this number of mock events, the accuracy is enough for our purpose.
Then, the probability is calculated by using Eq.~(\ref{KP-P-formula}).
We also checked the probability by direct Monte-Carlo simulation
for several cases and confirmed that the Eq.~(\ref{KP-P-formula})
is good enough.

\begin{figure}
\includegraphics[width=80mm]{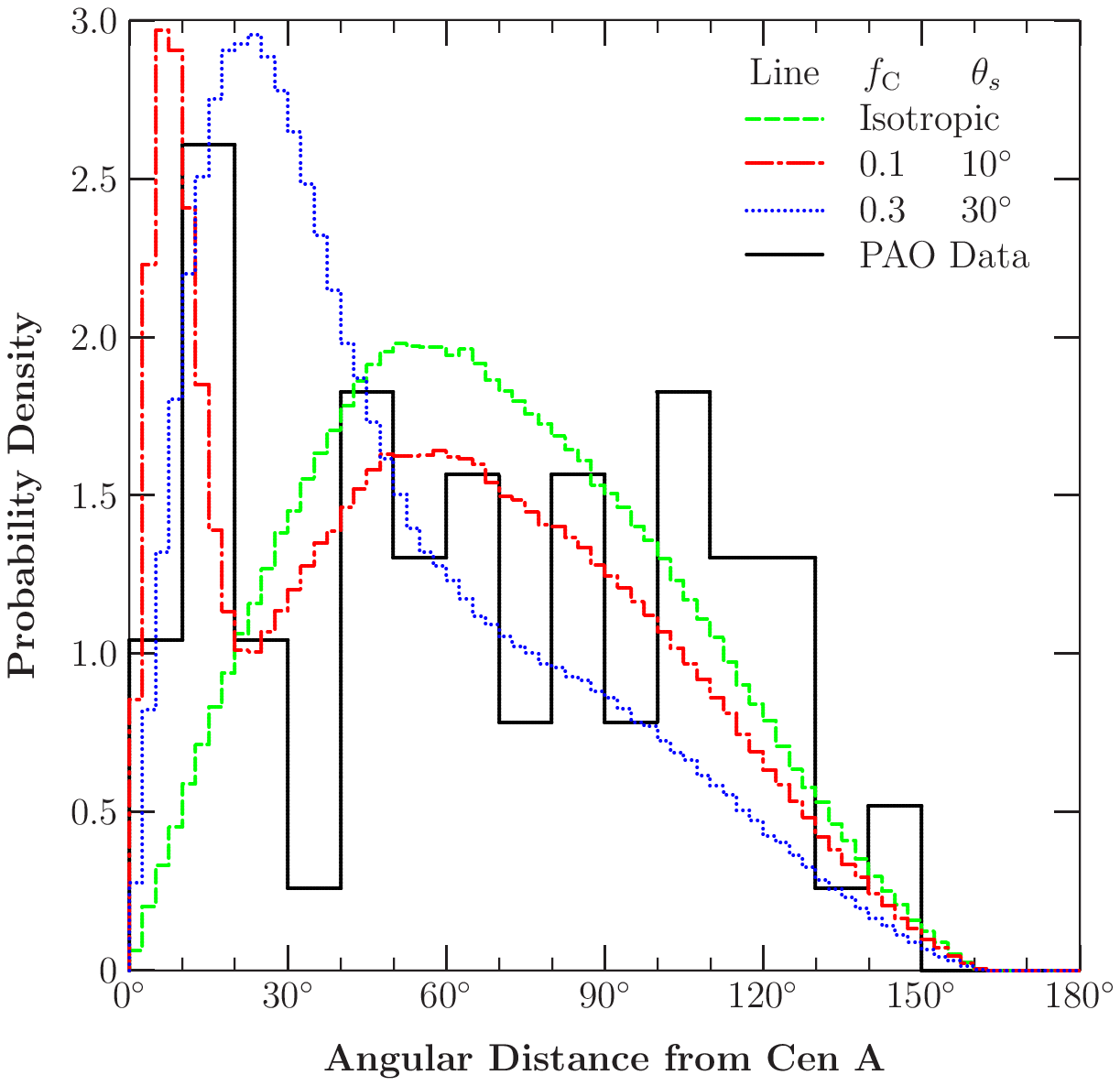}
\includegraphics[width=80mm]{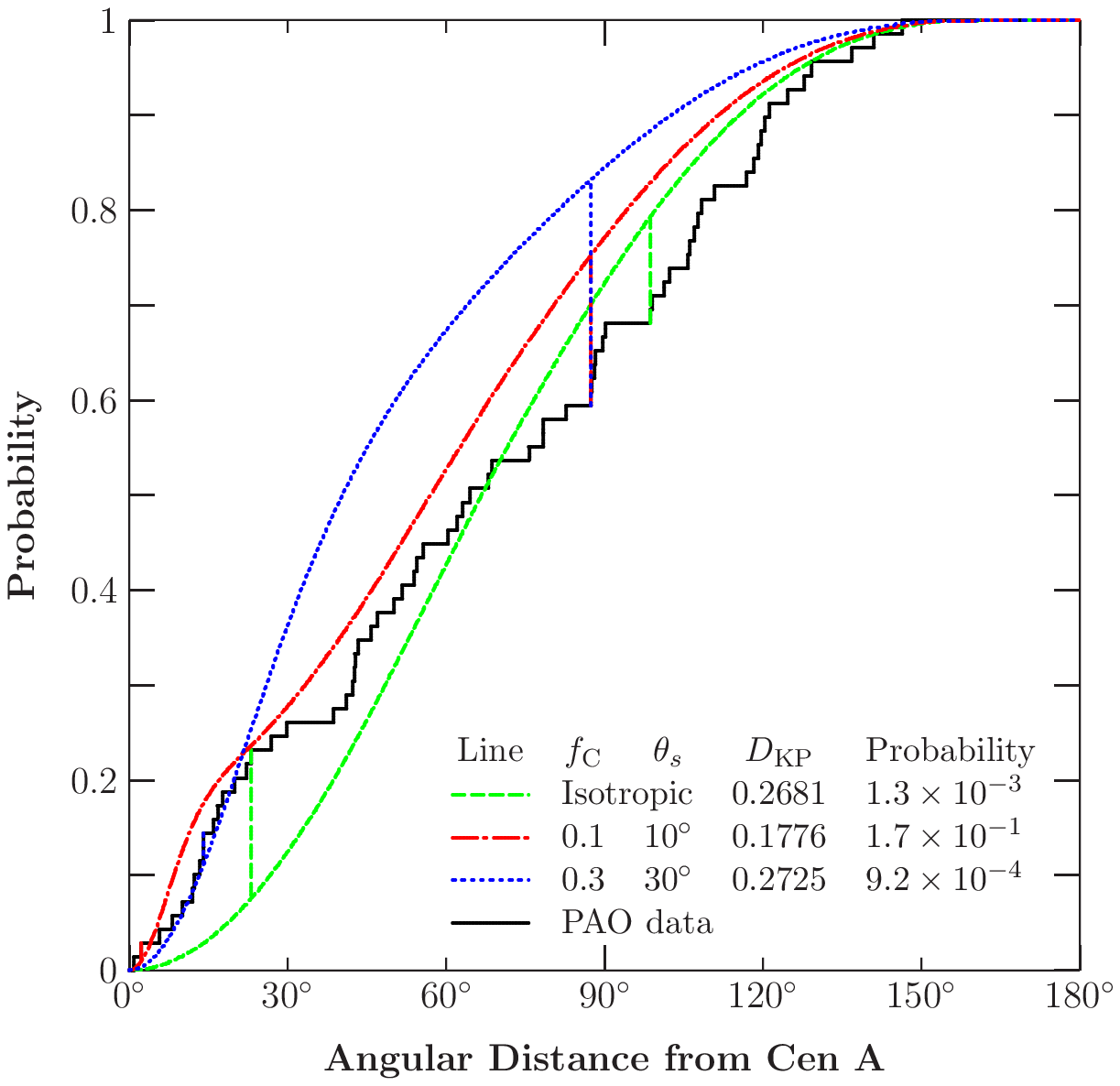}
\caption{CADDs (left panel) and their CPDs (right panel) of the PAO data,
the isotropic distribution, and the Cen A dominance model with the
parameter sets ($f_{\rm C}=0.1$, $\theta_s=10^\circ$) and
($f_{\rm C}=0.3$, $\theta_s=30^\circ$).
Vertical lines in the right panel represent the sizes and the locations of
the maximum differences between the CPD of the PAO data and those of
the models considered.}
\label{cpd-cadd}
\end{figure}

Fig.~\ref{cpd-cadd} shows CADDs and their CPDs of the PAO data,
the isotropic distribution, and two cases of the Cen A dominance model.
Close examination of CADD and its CPD is quite useful for understanding
the results of statistical analysis and what causes the discrepancy
between the data and the prediction of the model.
For a single source, CADD is simply the distribution of angular distances
of all UHECR from the source.
The CADD of the isotropic distribution has a bell shape, reflecting the
relative location of Cen A and the exposure function of the PAO experiment.
On the contrary,
the CADD of the PAO data has a distinguished peak at small angles,
which means that there is an excess of observed UHECR near Cen A
compared to the isotropic distribution.
The KP test on CADD indicates that the probability that this excess
of the PAO data is obtained from the isotropic distribution by chance
is $1.3\times10^{-3}$.
Thus, this excess can be attributed to the Cen A contribution.

\begin{figure}
\includegraphics[width=80mm]{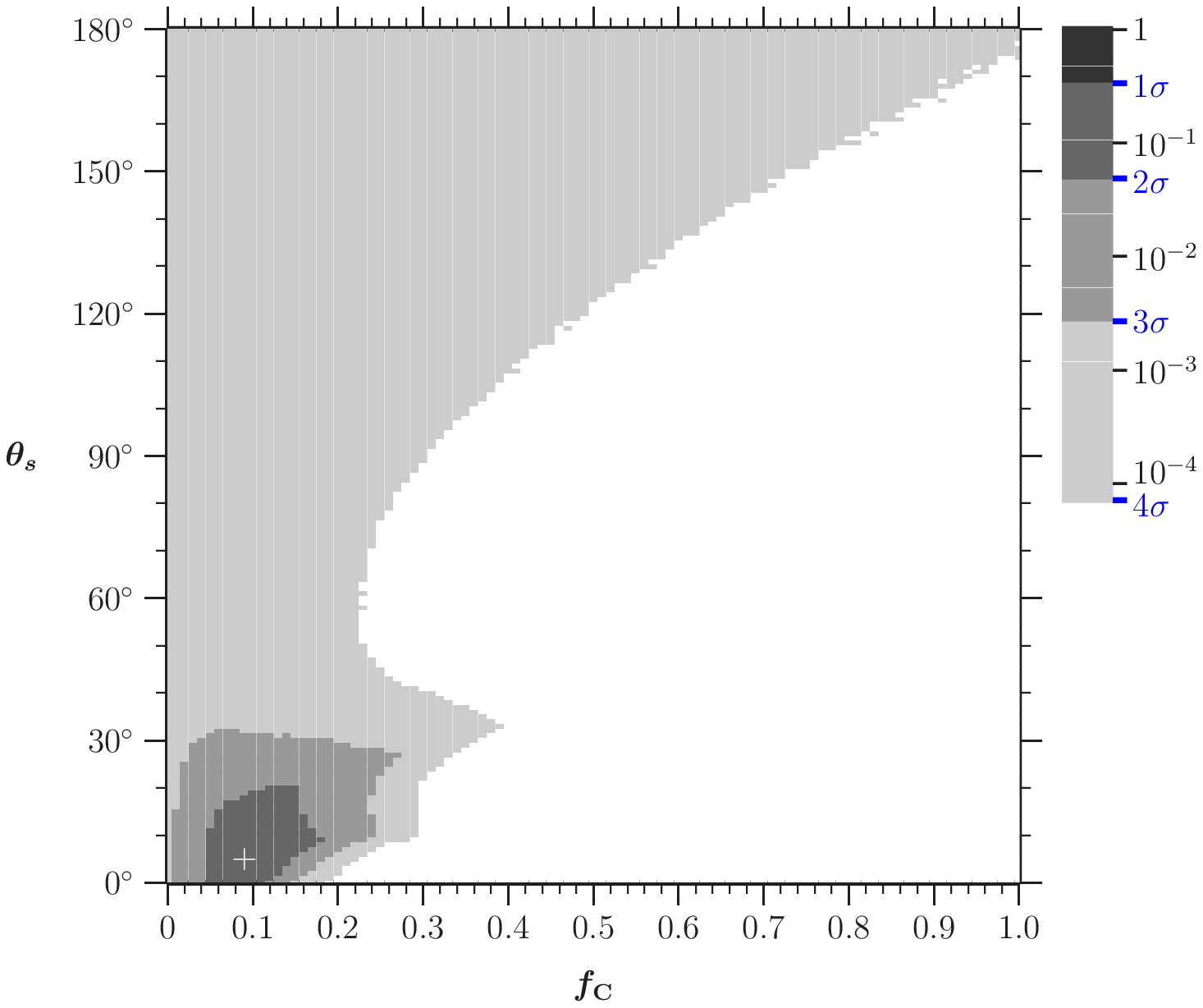}\
\includegraphics[width=80mm]{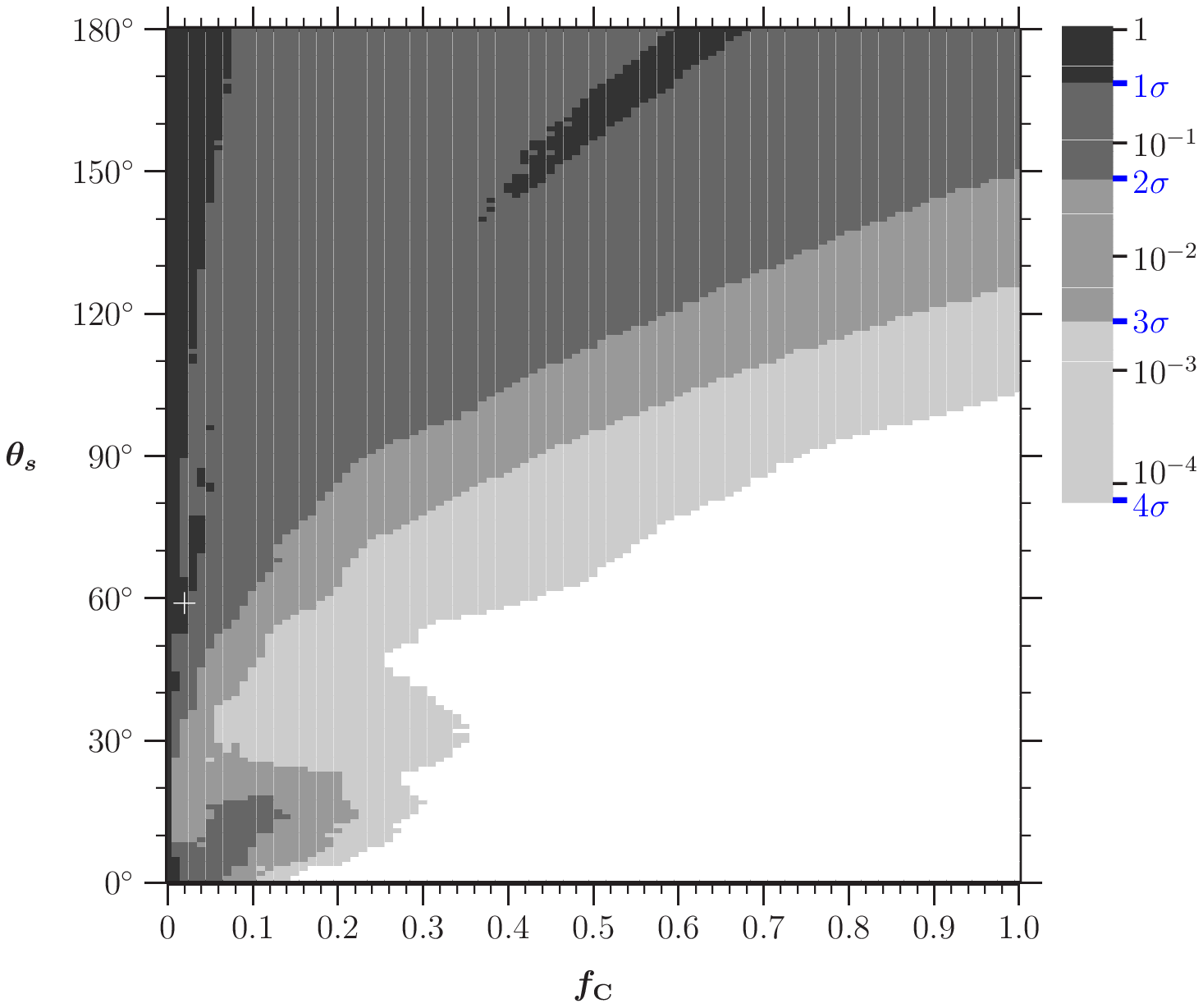}
\caption{Cen A fraction ($f_C$) and smearing angle ($\theta_s$) dependence of
probabilities that CADD (left panel) and FEVD (right panel) of the PAO data
are obtained from the Cen A dominance model.
The white $+$ marks the parameter values of the maximum probability.}
\label{prob-w-r}
\end{figure}

The Cen A dominance model has two parameters,
the Cen A fraction $f_{\rm C}$ and the smearing angle $\theta_s$,
which can be used to fit this excess peak and the remaining bell-shaped body.
The Cen A fraction $f_{\rm C}$ affects the height of the peak
at small angles relative to the bell-shaped body at larger angles.
The smearing angle $\theta_s$ changes the position and width of the peak.
As $\theta_s$ increases, the Cen A contribution tends to be isotropic
and the peak merges into the bell-shaped body.
Fig.~\ref{prob-w-r} shows the probability that CADD or FEVD of the observed
PAO data is obtained from the Cen A dominance model, as a function of two
parameters $f_{\rm C}$ and $\theta_s$.
Compared to the CADD method, the FEVD method seems to becomes rather
insensitive for small Cen A fraction or large smearing angles
where the distribution tends to be isotropic.
At angles smaller than $30^\circ$,
both the CADD and the FEVD methods give the similar and consistent results.
Thus, we mainly use the results of the CADD method to extract our conclusions.
For the CADD method, the best fit parameters at which the probability
reaches its maximum value are $f_{\rm C}\approx0.1$ and $\theta_s=5^\circ$,
giving the maximum probability $P_{\rm max}=0.29$.
We don't have the $1\sigma$ band ($P\ge0.32$) in the whole parameter ranges.
However, this is not so disappointing,
considering the simplicity of the Cen A dominance model that
the distribution of UHECR which are not contributed by Cen A
is assumed to be isotropic.
As you can notice from the CPDs in Fig.~\ref{cpd-cadd},
the parameters in the $2\sigma$ band fit the CPD at small angles well, but
fitting gets poor at large angles where the isotropic contribution dominates.
This makes the over-all fitting a little bit poor.
The main cause of this poor fit is the big dip observed at angles
$30^\circ$--$40^\circ$, as seen in the CADD of the PAO data
in Fig.~\ref{cpd-cadd}.  The big dip means the void region of UHECR
at angular distance $30^\circ$--$40^\circ$ from Cen A,
which is actually observed at the lower right region near the south pole
in the skymap of Fig.~\ref{skymap-data}.
This departure from isotropy due to the void region makes the assumption
of isotropic background in the Cen A dominance model a little bit poor one,
though it is still a valid first approximation.
Actually, this makes the best fit value of the smearing angle
smaller than the naively (or more appropriately) expected value
$10^\circ$--$20^\circ$ at which the peak of CADD is observed.
Considering this uncertainty concerning the assumption of isotropic background,
we extend the viable parameter range for the Cen A dominance model
to the parameter range of $2\sigma$ band ($P\ge0.046$):
$0.05\lesssim f_{\rm C}\lesssim 0.15$ and
$0^\circ\lesssim \theta_s\lesssim 20^\circ$.
In Section~\ref{sec2}, we mentioned that $f_{\rm C}$ is different from
$f_{\rm C,PAO}$, the fraction of Cen A contribution
as observed at the PAO experiment.
Because Cen A is located at the central region of the PAO exposure,
$f_{\rm C,PAO}$ is large than $f_{\rm C}$.
$f_{\rm C,PAO}$ can be obtained as a function of $f_{\rm C}$ and $\theta_s$
when we do the Monte-Carlo simulation.
In terms of $f_{\rm C,PAO}$,
the best-fit value is $f_{\rm C,PAO}=0.15$ and
the $2\sigma$ band is $0.09\lesssim f_{\rm C,PAO}\lesssim 0.25$,
which means that among 69 UHECR observed by PAO, about 10
($2\sigma$ band is $6\sim17$) UHECR can be attributed to Cen A contribution.

\section{Discussion and Conclusion}
\label{sec5}

Let us discuss the implications of the results
obtained in the previous section.
First of all,
though we started with the hypothesis that Cen A is a dominant source of
UHECR, what we can actually prove through our statistical analysis is
that there is a strong source of UHECR in the direction of Cen A.
To confirm that the actual source is Cen A, other evidences,
e.g., from the acceleration mechanism, energetics, or the energy spectrum,
are needed.

As seen in the previous section,
the major source of uncertainty in our statistical analysis is
the assumption that the background contribution is isotropic.
We used the UHECR data with energies larger than $5.5\times10^{19}\,{\rm eV}$,
which is larger than the GZK cutoff.
Thus, their sources are believed to be mostly located within the GZK radius
$\sim 100\,{\rm Mpc}$, where the matter distribution is not isotropic.
One important issue in the UHECR arrival directions is
whether the UHECR sources trace the matter distribution.
In this regard, it is important to check the existence of the correlation
between the UHECR arrival directions and the matter distribution
within the GZK radius.  There have been several studies on this,
and the existence of correlation is not yet conclusive
\cite{Koers:2008ba,Cuesta:2009ap,Takami:2009bv}.
Thus, we stick to the isotropic background
instead of modeling the matter-tracing background.
Concerning the background contribution,
we again draw your attention to the fact that
the CADD of the PAO data depicted in Fig.~\ref{cpd-cadd} shows
the large deficit at $30^\circ$-$40^\circ$ bin and
another excess at $100^\circ$-$130^\circ$ range
compared to the isotropic distribution.
The deficit is due to the void region near the south pole in the PAO data.
The chance probability that this void region is obtained from the isotropic
distribution is about $5.0\times10^{-3}$ by the same CADD method and KP test.
Not only the excess in the Cen A direction but also this void region
makes the PAO data anisotropic.
Another excess may be due to another point sources.
We have examined the possibility of the existence of another point source
at the region of angular distance $100^\circ$--$130^\circ$ from Cen A,
but found that this excess is consistent with isotropy.

The importance of searching for the point sources of UHECR is that
it is the starting point of cosmic ray astronomy.
A known point source of UHECR can be used to probe the intergalactic
magnetic fields in the vicinity of the source.
For Cen A, this kind of study was done in Ref.~\cite{Yuksel:2012ee}.
Here, we provide the order estimate from our results for comparison.
Once we accept the results in the previous section that
the fraction $f_{\rm C,PAO}=0.15$ of the observed UHECR by PAO is contributed
by Cen A and their average deflection angle is around $10^\circ$,
we can estimate the rms magnitude of the magnetic fields
in the vicinity of Cen A.
If we assume that the intergalactic magnetic field is composed of
random patches of magnetic fields,
the rough estimate for the deflection angle is given by
\cite{Nagano:2000ve}
\begin{equation}
\delta\theta = 0.8^\circ\,Z
\left(\frac{E}{10^{20}\,{\rm eV}}\right)^{-1}
\left(\frac{d\ell_c}{10\,{\rm Mpc^2}}\right)^{1/2}
\left(\frac{B}{10^{-9}\,{\rm G}}\right),
\end{equation}
where $E$ is the energy of the cosmic ray particle,
$d$ is the size of the magnetic field extension,
$\ell_c$ is the average size of patches, and
$B$ is the magnetic field strength.
If we select 10 nearest UHECR from Cen A,
the average energy is $E\sim7.0\times10^{19}\,{\rm eV}$,
and the average defection angle is $\delta\theta\sim10^\circ$
(The maximum deflection angle is $\delta\theta\sim15^\circ$.).
Inserting these values, $Z=1$ (the proton),
and the distance $d\sim4\,{\rm Mpc}$,
we obtain an estimate $B\sim14\,(\ell_c/{\rm Mpc})^{-1/2}\,{\rm nG}$.
This gives the rough strength of the intergalactic
magnetic field in the vicinity of Cen A.

In conclusion, we examined the possibility
that Cen A is a dominant source of UHECR observed by PAO,
by using the statistical analysis of the arrival direction distribution.
We set up the Cen A dominance model for the UHECR sources,
in which Cen A contributes the fraction $f_{\rm C}$ of the whole UHECR
with energy above $5.5\times10^{19}\,{\rm eV}$ and
the isotropic background contributes the remaining $1-f_{\rm C}$ fraction.
The effect of the intergalactic magnetic fields on the bending
of the trajectory of UHECR originated from Cen A
is parameterized by the gaussian smearing angle $\theta_s$.
We adopted CADD and FEVD methods
for the reduction of the arrival direction distribution,
and the KP test to compare the observed and the expected CADD and FEVD
and to estimate the significance level of the similarity.
We observed the excess of UHECR in the Cen A direction in CADD.
Then we tried to fit the CADD of the PAO data by varying two parameters
$f_{\rm C}$ and $\theta_s$ of the Cen A dominance model.
The best-fit parameter values are
$f_{\rm C}\approx0.1$ ($f_{\rm C,PAO}\approx0.15$) and $\theta_s=5^\circ$
with the probability $P=0.29$.
Considering the uncertainty concerning the assumption of isotropic background
in the Cen A dominance model, we extend the viable parameter ranges to
the $2\sigma$ band,
$0.09\lesssim f_{\rm C,PAO}\lesssim 0.25$ and
$0^\circ\lesssim \theta_s\lesssim 20^\circ$.
It supports the existence of a point source in the direction of Cen A,
which is extended by the action of intergalactic magnetic fields.
If Cen A is actually the source responsible for the observed excess,
the intergalactic magnetic field in the vicinity of Cen A
is estimated to be $B\sim14\,(\ell_c/{\rm Mpc})^{-1/2}\,{\rm nG}$
from the average deflection angle of the excess UHECR.

\section*{ACKNOWLEDGMENT}

This research was supported by Basic Science Research Program
through the National Research Foundation (NRF) funded by
the Ministry of Education, Science and Technology (2011-0002617).

\end{document}